\documentclass[showpacs,twocolumn,pre,floatfix]{revtex4}
\usepackage{graphicx}

\newcommand{\latin}[1]{\textit{#1}}
\newcommand{\etal}{\latin{et al}}

\newcommand{\eqref}[1]{(\ref{#1})}

\begin{document}

\title{Late stage kinetics for various wicking and spreading problems}

\author{Patrick B. Warren}
\affiliation{Unilever R\&D Port Sunlight, 
Bebington, Wirral, CH63 3JW, UK.}

\date{Oct 13, 2003}

\begin{abstract}
The kinetics of spreading of a liquid drop in a wedge or V-shaped
groove, in a network of such grooves, and on a hydrophilic strip, is
re-examined.  The length of a droplet of volume $\Omega$ spreading in
a wedge after a time $t$ is predicted to scale as $\Omega^{1/5}
t^{2/5}$, and the height profile is predicted to be a parabola in the
distance along the wedge.  If the droplet is spreading radially in a
sparse network of V-shaped grooves on a surface, the radius is
predicted to scale as $\Omega^{1/6} t^{1/3}$, provided the liquid is
completely contained within the grooves.  A number of other results
are also obtained.
\end{abstract}


\maketitle

\section{Introduction}
Wetting in complex geometries and on rough surfaces provides a wealth
of fascinating non-linear hydrodynamics problems, as well as being of
commercial importance in numerous industrial sectors.  Perhaps the
first kind of problem to be considered was the penetration of liquid
into porous materials, where Washburn in 1921 demonstrated that the
distance attained by the wetting front follows a $t^{1/2}$ law where
$t$ is time \cite{Washburn}.  Much later the spreading of droplets on
flat surfaces was addressed by various workers, such as Tanner
\cite{Tanner} and Lopez \etal\ \cite{LMR}, although it took some time
for the subtleties of the physics at the wetting front to be resolved
\cite{deGennes,LJ,BB,ODB}.  Generally, the wetting front advances with
a $t^{\alpha}$ law where $\alpha$ is a small exponent which depends on
the geometry of spreading and the origin of the driving force.  For
example, $\alpha=1/10$ for a drop spreading radially driven by surface
tension (Tanner's law), and $\alpha=1/8$ for a droplet spreading
radially driven by gravity (see Oron \etal\ \cite{ODB} for a summary
of results).

The kinetics of wetting on rough surfaces has also been investigated
experimentally and theoretically \cite{CCS,GCST,RYOT}.  A paradigm for
this problem is the spreading of a liquid in a wedge or V-shaped
groove \cite{MRRY,RY,GCS}; indeed wetting in a network of V-shaped
grooves has been invoked recently for oil spreading on skin
\cite{DAL}.  Another kind of problem that has been considered is the
wetting of hydrophilic strips \cite{DTR}, as an example of
wetting in a controlled microstructure that might be contemplated in a
microfluidic device.  In all these problems, a $t^{1/2}$ spreading law
has been observed, but in the cases considered thus far, there has
been a reservoir which provides liquid at essentially a constant
pressure.  In the present paper, the problems of spreading in a wedge,
in a network of V-shaped grooves, and on a hydrophilic strip are
revisited.  It is found that in the absence of a reservoir, the
spreading law changes to $t^{\alpha}$ with $\alpha<1/2$, similar to
Tanner's law and related problems.

These problems are first approached by scaling arguments developed in
the next section.  The \latin{bona fides} of the scaling arguments is
established by rederiving some known results for spreading on flat
surfaces.  In a further section, the scaling exponents are recovered
by similarity analysis on the underlying partial differential
equations which govern spreading.  This also allows the scaling shape
of the spreading drops to be computed.

\section{Scaling arguments}
\begin{figure}
\includegraphics{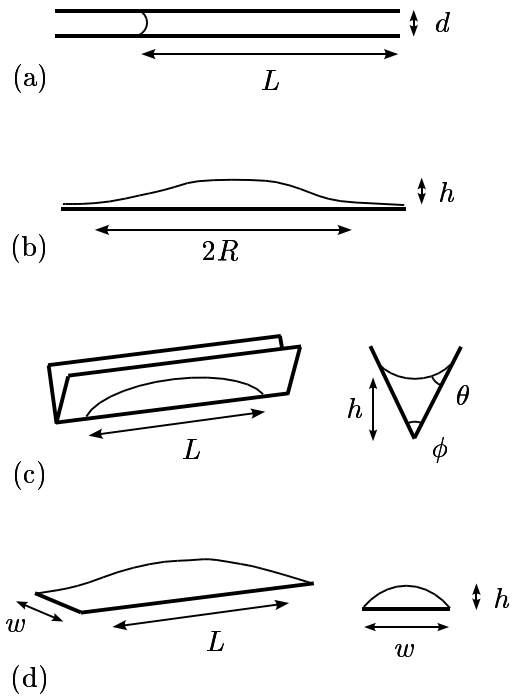}
\caption[?]{Various wetting problems: (a) wicking into a capillary,
(b) spreading on a flat substrate, (c) spreading in a wedge, and
(d) spreading along a hydrophilic strip.\label{fig:diagram}}
\end{figure}

The Washburn problem of a liquid being drawn into a capillary tube of
internal dimension $d$ shown in Fig.~\ref{fig:diagram}(a) is
considered first \cite{Washburn}.  This is a model for penetration of
liquid into a porous material for which $d$ interpreted as a mean pore
size.  The arguments here are very familiar, but form the basis for
the more complex problems considered below.

Once the liquid has penetrated a sufficient distance $L\gg d$, a
Poiseuille law obtains for the liquid velocity and the penetration
rate, thus
\begin{equation}
\frac{dL}{dt}\sim\frac{d^2}{\eta}\,\frac{\Delta p}{L}\label{wash1}
\end{equation}
where $\eta$ is viscosity, and the pressure drop
\begin{equation}
\Delta p\sim\sigma/d\label{wash2}
\end{equation}
is due to the surface tension $\sigma$ of the curved surface at the
wetting front, at a mean curvature $\sim 1/d$.  All geometric factors
associated with the shape of the tube and a finite contact angle have
been dropped, although a contact angle $\theta<\pi/2$ is required for
imbibition to take place.  Combining Eqs.~\eqref{wash1} and
\eqref{wash2} gives
\begin{equation}
\frac{dL}{dt}\sim\frac{\sigma}{\eta}\,\frac{d}{L}\label{wash3}
\end{equation}
which integrates to
\begin{equation}
L\sim(\sigma t d/\eta)^{1/2}.\label{wash4}
\end{equation}
This is the simplest form of the Washburn equation \cite{Washburn}.
The result arises from a constant pressure drop acting over an
increasing length of liquid, which responds by flowing according to
the Poiseuille law.  As we shall see below, this can be the case for
many situations where a reservoir of liquid is present, but if a
reservoir is absent, the rate of spreading can be much slower.

Next, the problem of a drop of liquid spreading on a flat surface is
considered, as shown in Fig.~\ref{fig:diagram}(b).  Usually the
problem is approached by an appeal to the hydrodynamics in the
vicinity of the moving contact line \cite{deGennes,LJ}, but it can be
analysed using similar concepts to the Washburn problem.  Whilst only
previously known results are recovered, the approach serves to
illustrate further the arguments that will be used for the other
problems.

Consider a drop of liquid spreading on a flat surface, in the case of
complete wetting.  Let a measure of the radius of the spreading drop
be $R$ and the height in the centre be $h$.  In the lubrication
approximation, assuming a scaling shape of the droplet, all velocities
will be proportional to $(h^2\!/\eta)(\Delta p/R)$ (compare Poiseuille
law above) where $\Delta p$ is the pressure drop between the centre
and the radius $R$.  In particular the drop radius is expanding at a
rate
\begin{equation}
\frac{dR}{dt}\sim\frac{h^2}{\eta}\,\frac{\Delta p}{R}.
\end{equation}
First consider the capillary spreading case where the pressure
gradient is due to surface tension $\sigma$.  Simple geometry shows
that the mean curvature at the centre of the droplet for $h\ll R$ is
$\sim h/R^2$ therefore the pressure drop is
\begin{equation}
\Delta p\sim\sigma h/R^2
\end{equation}
and hence
\begin{equation}
\frac{dR}{dt}\sim\frac{\sigma}{\eta}\,\frac{h^3}{R^3}.
\end{equation}
If $h$ were to be constant, as in the Washburn problem, this would be
enough to determine the spreading rate.  Here, though, a second
relation connecting $R$ and $h$ must be sought.  The desired relation
follows from the conservation of drop volume $\Omega$,
\begin{equation}
h R^2\sim\Omega.
\end{equation}
Thus
\begin{equation}
\frac{dR}{dt}\sim\frac{\sigma}{\eta}\,\frac{\Omega^3}{R^9}
\end{equation}
which integrates to 
\begin{equation}
R\sim (\sigma t\Omega^3\!/\eta)^{1/10}.
\end{equation}
This result is Tanner's law \cite{Tanner}.  The basic scaling $R\sim
\Omega^{3/10}\,t^{1/10}$ is well documented and has been
experimentally verified \cite{deGennes}.

For the case where the spreading is driven by gravity, one has $\Delta
p\sim\rho g h$ where $\rho$ is the mass density and $g$ is the
acceleration due to gravity.  Following the same line of argument as
above, one obtains $R\sim (\rho g t\Omega^3\!/\eta)^{1/8}$ \cite{LMR}.
The behaviour crosses over from capillary spreading to gravity
spreading when the Bond number $\rho g R^2\!/\sigma$ increases.  Since
$R$ is increasing, this means that capillary spreading always crosses
over to gravity spreading if one waits long enough.  The weak increase
in spreading rate has been observed experimentally \cite{CCS}.

Another case that can be considered is planar or one-dimensional
spreading.  The only thing which changes is the volume conservation
law which becomes $hL\sim\Omega$ where $L$ replaces $R$ as the
measure of extent of spreading, and $\Omega$ is a volume per unit
length.  This yields $L\sim (\sigma t\Omega^3\!/\eta)^{1/7}$ and $L\sim
(\rho g t\Omega^3\!/\eta)^{1/5}$ for capillary \cite{Tanner} and gravity
\cite{LMR} spreading respectively.

In the next problem, exactly analogous arguments are applied to the
case of spreading in a wedge, shown in Fig.~\ref{fig:diagram}(c).  In
the case of spreading from a reservoir, this problem has been
addressed by Romero and Yost \cite{RY}.  The basic idea is that one
has scale invariance, with the depth $h$ of fluid being the only
relevant length scale.  Hence the transverse curvature of the
interface $\propto 1/h$. Thus, provided the droplet has become
sufficiently extended so that the contribution of the longuitudinal
curvature to the mean curvature can be neglected, the pressure
$p\propto (-)\sigma/h$ where the negative sign obtains if the surface
is convex into the liquid.  This is the case if $2\theta+\phi<\pi$
where $\theta$ is the contact angle and $\phi$ is the wedge angle as
in Fig.~\ref{fig:diagram}(c).  In this case, the pressure becomes more
negative as the amount of fluid is in the wedge gets smaller.  This
provides a pressure gradient which drives the liquid from regions of
high loading to low loading.

Even though the liquid has a free surface, a Poiseuille-like law
obtains,
\begin{equation}
\frac{dL}{dt}\sim\frac{h^2}{\eta}\,\frac{\Delta p}{L}
\end{equation}
where $L$ is a measure of the extent of spread of the liquid drop.  
The pressure drop follows from the above arguments, $\Delta
p\sim\sigma/h$, and therefore
\begin{equation}
\frac{dL}{dt}\sim\frac{\sigma}{\eta}\,\frac{h}{L}\label{wedge2}
\end{equation}
in perfect analogy to Eq.~\eqref{wash3} in the Washburn problem.  The
difference here is that $h$ is a dynamic variable which, similar to
the Tanner's law derivation above, is found by a volume conservation
law.  In this case $AL\sim\Omega$ where $A\sim h^2$ is the cross
section area occupied by the liquid, and $\Omega$ is the volume of
liquid.  Thus $h\sim (\Omega/L)^{1/2}$. Substituting into
Eq.~\eqref{wedge2} results in
\begin{equation}
\frac{dL}{dt}\sim\frac{\sigma}{\eta}\,\frac{\Omega^{1/2}}{L^{3/2}}
\end{equation}
which integrates to 
\begin{equation}
L\sim(\sigma t\Omega^{1/2}\!/\eta)^{2/5}.\label{wedge3}
\end{equation}
This spreading law, $L\sim\Omega^{1/5}t^{2/5}$, is a new prediction.
It can also be recovered by a more detailed analysis of the underlying
partial diffential equations which will be given in a later section.

Note that if the liquid had been spreading along the wedge from a
reservoir, this would correspond to $h\sim\mathrm{constant}$.  This
leads to the Washburn result $L\sim t^{1/2}$ and is the origin of the
scaling law obtained by previous workers \cite{MRRY,RY}.

The next case to be considered is the problem of a droplet spreading
into a network of grooves.  This has also been considered by various
groups \cite{CCS,GCST,DAL}.  If spreading occurs from a reservoir,
then the front advances with a $t^{1/2}$ Washburn-like law.  However
the case where the liquid is completely confined in the grooves is
different.  The scaling law in this case follows from arguments
similar to those already applied to a drop spreading in a wedge.  The
analysis assumes that the grooves are rather sparse on the surface, in
particular that the volume of liquid in the junction zones can be
neglected compared to the volume contained in the grooves.

Consider therefore a drop of liquid spreading in a sparse random
network of grooves.  It will spread essentially radially.  Suppose
that the groove cross section is a V-shape, so that both the capillary
pressure and the Poiseuille scaling laws can be taken over from the
case of spreading in a wedge.  The fact that the grooves are randomly
inclined with respect to the radial pressure gradient only introduces
an additional numerical prefactor \cite{DAL}.  Thus Eq.~\eqref{wedge2}
above still holds (with $L$ replaced by $R$).  What changes is the
volume conservation law: as $R$ increases, more grooves become filled.
If the length of grooves per unit area is $l^{-1}$, where $l$ is a
characteristic groove spacing on the surface, the total length of
grooves occupied by the liquid $\sim R^2\!/l$ and the volume $\Omega\sim
h^2R^2\!/l$.  Eliminating $h$ between this and Eq.~\eqref{wedge2} (with
$L$ replaced by $R$) gives $dR/dt\sim (\sigma/\eta)
(l\Omega)^{1/2}\!/R^2$.  This integrates to
\begin{equation}
R\sim (\sigma t(l\Omega)^{1/2}\!/\eta)^{1/3}.\label{eqnet}
\end{equation}
The basic prediction therefore is that the spreading rate should slow
from the $t^{1/2}$ Washburn-like law for spreading from a reservoir,
to an $R\sim\Omega^{1/6}t^{1/3}$ law as the liquid becomes confined to
the grooves.  The result is confirmed by a more detailed analysis of
the underlying partial differential equations given later.

The final problem that is considered is spreading along a hydrophilic
strip, shown in Fig.~\ref{fig:diagram}(d).  The case where
spreading occurs from a reservoir has been investigated both
theoretically and experimentally, and it is found that the spreading
front advances with a Washburn-like $t^{1/2}$ law \cite{DTR}.  The
situation in the absence of a reservoir was not investigated though.
Once again, the absence of a reservoir leads to a slower rate of
spreading, and the scaling law can be determined using analogous
arguments to those applied above.

Consider a drop of liquid which is completely wetting on a hydrophilic
strip of width $w$.  In the late stages, let the height of the
liquid above the surface be $h\ll w$, and a measure of the extent of
spreading be $L\gg w$.  The capillary pressure which drives the
spreading is due to the transverse curvature of the interface, which
for $h\ll w$ is $\sim h/w^2$.  The pressure gradient is thus $\Delta
p/L\sim\sigma h/w^2L$.  In the lubrication approximation with $h\ll
w$, a Poiseuille-like law obtains for the fluid velocity with $h$
being the relevant length scale.  Hence the rate of extension of the
droplet obeys $dL/dt \sim (h^2\!/\eta)(\Delta p/L) \sim
(\sigma/\eta)h^3\!/w^2L$.  Volume conservation indicates $hwL\sim\Omega$
and eliminating $h$ between this and the above spreading rate finds
$dL/dt \sim (\sigma/\eta) \Omega^3\!/w^5L^4$.  This integrates to
$L\sim(\sigma t\Omega^3\!/\eta w^5)^{1/5}$.  This is the predicted
scaling law for the late stages of spreading in this problem, in the
absence of a reservoir.

\begin{figure}
\includegraphics{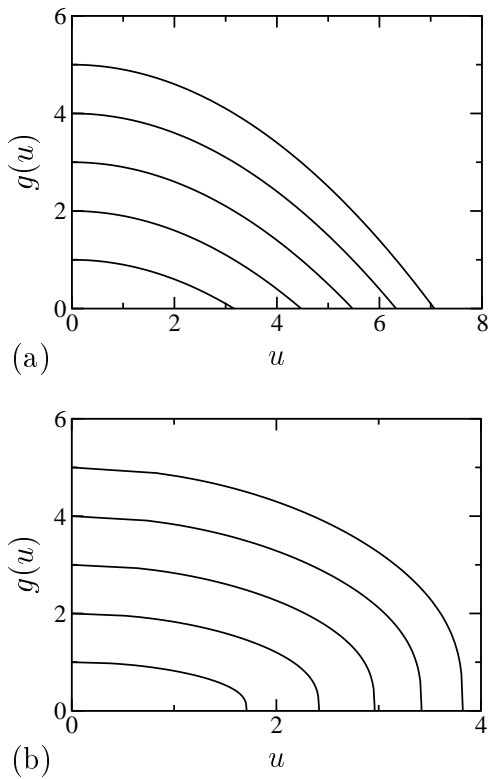}
\caption[?]{Shapes of droplets spreading in (a) a wedge and (b) a
network of grooves.  These are similarity solutions $g(u)$ found by
integrating Eq.~\eqref{odeeq} with boundary conditions $g = A$ and $g'
= 0$ at $u = 0$, for $A=1(1)5$, where $u$ is a scaled distance.  In
case (a) the shape is a parabola given by Eq.~\eqref{eqpara}.  In case
(b) the shape is obtained by numerical integration.\label{fig:shape}}
\end{figure}

\section{Similarity methods}
The results obtained above can also be derived by using similarity
methods to analyse the underlying partial differential equations (see
for example Ref.~\cite{BB}).  Focus first on the problem of spreading
in a wedge shown in Fig.~\ref{fig:diagram}(c).  An equation which
expresses local conservation of liquid in the wedge is
\begin{equation}
\frac{\partial A}{\partial t} +
\frac{\partial(A\overline v)}{\partial x} = 0\label{areaeq}
\end{equation}
where $A\sim h^2$ is the local cross section area occupied by liquid,
$\overline v$ the mean velocity of the liquid, and $x$ is distance
along the wedge.  The Poiseuille law indicates that the mean velocity
follows
\begin{equation}
\overline v\propto -\frac{h^2}{\eta}\frac{\partial p}{\partial x}.
\end{equation}
The arguments above show that $p\propto-\sigma/h$ thus
\begin{equation}
\frac{\partial p}{\partial x}\propto
\frac{\sigma}{h^2}\frac{\partial h}{\partial x}.\label{gradpeq}
\end{equation}
Combining Eqs.~\eqref{areaeq}--\eqref{gradpeq} gives the following
equation for the time evolution of the depth of liquid in the groove
(compare Eqs.~(8a)--(8c) of Romero and Yost \cite{RY})
\begin{equation}
\frac{\partial(h^2)}{\partial t} = 
K\frac{\sigma}{\eta} \frac{\partial}{\partial x}\Bigl(
h^2\frac{\partial h}{\partial x}\Bigr).\label{pdeeq}
\end{equation}
The dimensionless coefficient $K(\theta,\phi)$ is given by Eq.~(8c) in
Romero and Yost \cite{RY} in terms of the static contact angle
$\theta$ and the wedge angle $\phi$.  For spreading to occur, $K>0$ is
required.  This corresponds to $2\theta+\phi<\pi$, or a liquid
interface which is convex into the liquid.  For the remainder of the
discussion, the factor $K\sigma/\eta$ and other trivial numerical
prefactors can be adsorbed into the units of time, and will be
omitted.

Eq.~\eqref{pdeeq} is basically a non-linear diffusion equation and one
can seek similarity solutions of the form
\begin{equation}
h(x,t) \sim t^{-\beta} \, g(x t^{-\alpha})\label{simeq}
\end{equation}
where $u = x t^{-\alpha}$ is the similarity variable and $\alpha$ is
the exponent in the spreading law.  Substituting this in
Eq.~\eqref{pdeeq} obtains both an exponent relation
\begin{equation}
2\alpha+\beta=1\label{expnt1}
\end{equation}
which must be satisfied for the similarity solution to hold, and an
ordinary differential equation (ODE) for the similarity function
\begin{equation}
gg''+2(g')^2+\alpha u g'+\beta g=0.\label{odeeq}
\end{equation}
This is a non-linear second order ODE with boundary conditions
$g(0)=A$ (which is set by the drop volume) and $g'(0)=0$ (required by
symmetry).

A second exponent relation follows from the integrated conservation
law.  Volume conservation dictates that
$\Omega\propto\int_{-\infty}^\infty h^2\,dx$ is constant.  Inserting
the similarity solution in this shows that $\Omega\sim
t^{\alpha-2\beta}\times \int_{-\infty}^\infty g^2\,du$ is constant,
and therefore
\begin{equation}
\alpha=2\beta.\label{expnt2}
\end{equation}
Solving Eqs.~\eqref{expnt1} and \eqref{expnt2} gives
\begin{equation}
\alpha=2/5,\quad\beta=1/5.\label{sngleq}
\end{equation}
Thus the spreading law $L\sim t^{2/5}$ of the preceeding section is
recovered.  The dependence on $\Omega$ and $\sigma/\eta$ can be
determined by dimensional analysis.

To complete the discussion, the ODE for the similarity function $g(u)$
can be solved.  Inserting Eq.~\eqref{sngleq} into Eq.~\eqref{odeeq}
results in $gg'' + 2(g')^2 + 2u g'\!/5 + g/5=0$, with $g(0)=A$ and
$g'(0)=0$.  Remarkably, this equation has an extremely simple closed
form solution,
\begin{equation}
g = A - u^2\!/10.\label{eqpara}
\end{equation}
Note that $g \to 0$ for $u \to u_0 = \sqrt{10 A}$, so the spreading
drop in the groove has a finite extent.  Some examples of the shape
for different values of $A$ are shown in Fig.~\ref{fig:shape}(a).  The
basic prediction is that the height profile of the liquid surface for
a droplet spreading in wedge is a parabola in the distance along the
wedge.  To be specific, from Eqs.~\eqref{simeq}, \eqref{sngleq} and
\eqref{eqpara}, 
\begin{equation}
h(x,t)=h_0\,[1-(x/x_0)^2],\quad (|x|<x_0)
\end{equation}
where the height $h_0\sim t^{-1/5}$ and the maximum extent of
spreading $x_0\sim t^{2/5}$.

For the case of spreading in a network of grooves,
Eqs.~\eqref{pdeeq}--\eqref{odeeq} remain the same (apart from
numerical prefactors \cite{DAL}), but the integrated conservation law
changes to $\Omega\propto\int_0^\infty h^2\,2\pi r\,dr/l$ (where $r$
should replace $x$ in Eq.~\eqref{pdeeq}).  Substituting the similarity
solution shows that $\alpha=\beta$ must hold in this case.  Combining
this with Eq.~\eqref{expnt1} gives $\alpha=\beta=1/3$, thus the
previous scaling exponent is recovered.  Unfortunately the
corresponding ODE does not have a closed form solution.  Some examples
of shapes obtained by numerical integration are shown in
Fig.~\ref{fig:shape}(b).  Again $g \to 0$ for $u \to u_0$ but in this
case $g$ has a singularity: $g\sim (u_0-u)^{1/3}$ as $u\to u_0$ from
below.

Finally, the problem of a liquid spreading along a hydrophilic strip
shown in Fig.~\ref{fig:diagram}(d) is discussed.  The governing
partial differential equation has been obtained by Darhuber \etal\ for
this problem \cite{DTR}.  It is ${\partial h}/{\partial t} \sim
\partial[h^3 ({\partial h}/{\partial x})]/{\partial x}$ where the
prefactor can be found from Eq.~(7) in Ref.~\cite{DTR}.  Substituting
the similarity trial solution Eq.~\eqref{simeq} in this yields an ODE
$g^3g'' + 3 g^2 (g')^2 + \alpha u g' + \beta g=0$\ with $g(0)=A$ and
$g'(0)=0$ being initial conditions, and an exponent relation
$2\alpha+3\beta=1$.  It follows from
$\Omega\propto\int_{-\infty}^\infty w h\,dx$ that $\alpha=\beta$.
Solving this together with the preceeding exponent relation gives
$\alpha=\beta=1/5$, thus recovering the exponent of the preceeding
section.  Numerical integration of the ODE gives results which
resemble those for the case of spreading in a network of grooves,
shown in Fig.~\ref{fig:shape}(b).  Again $g\to 0$ as $u\to u_0$ with
$g\sim(u_0-u)^{1/3}$ as $u\to u_0$ from below (compare Fig.~3 of
Ref.~\onlinecite{DTR}).

In all these cases, the fact that $g\to 0$ for $u\to u_0$ stands in
contrast to the similarity solution for Tanner's law for which there
is no point where $g\to 0$ \cite{BB}.  In that case one has to invoke
an additional microscopic mechanism to account for the shape of the
edge of the drop.

The case of spreading from a reservoir can also be treated with
variant of the above analysis; in fact the essential arguments are
already given by Romero and Yost \cite{RY}, and Darhuber \etal\
\cite{DTR}.  In the case of a reservoir, $h$ is constant at the
reservoir edge which we define to be the point $x=0$.  In terms of the
similarity solution Eq.~\eqref{simeq}, this forces $\beta=0$.
Combining this with the exponent relation in Eq.~\eqref{expnt1} or the
analogous exponent relation for the strip problem shows
$\alpha=1/2$ for all cases.  Thus the Washburn-like spreading law is
recovered for spreading from a reservoir, independent of whether
spreading takes place in a wedge, in a network of grooves, or along a
strip.

\section{Discussion}
The main results concern the kinetics of spreading in various
geometries.  New predictions are made for the scaling laws governing
the rate at which a droplet spreads in a wedge or V-shaped groove, in
a network of such grooves, and on a hydrophilic strip.  These
are established both by simple scaling arguments and by similarity
solutions of the underlying partial differential equations.  The
asymptotic shapes of the spreading droplets have also been considered.

Previous work on these problems has assumed the presence of a
reservoir which supplies liquid at constant pressure.  This results in
spreading laws which are essentially the same as the Washburn law for
penetration of a liquid into a porous material, or into a capillary.
The analysis here complements this previous work by considering the
problems in the absence of a reservoir.  This will apply in the late
stages of wetting when the reservoir becomes exhausted, or if there is
only a small amount of liquid present.

These predictions could be tested by simulation or in experiments.
The case of a droplet in a wedge would seem to be particularly simple,
for instance a drop will spread in a right-angled corner provided the
contact angle is less than $45^\circ$.  The prediction that the
scaling shape of the droplet should be a parabola should also be
tested.

I thank Alex Lips for discussions and encouragement.


\end{document}